\documentclass[12pt]{iopart}

\def\be{\begin{equation}}
\def\ee{\end{equation}}

\def\sup{\Sigma}
\def\suup{\sigma}

\def\mm2{{\cal V}}

\def\embed{\chi}
\def\d{{\mbox d}}
\def\g{g}
\newcommand{\bm}[1]{\mbox{\boldmath $#1$}}
\def\spI{(\mm2_I, g_I)}
\def\spE{(\mm2_E, g_E)}
\def\ww{W_2}
\def\www{W_3}

\def\Journal#1#2#3#4#5#6{#5 {#1} {\bf #2} #3}

\def\PRD{\em Phys. Rev. D }
\def\GRG{\em Gen. Rel. Grav.}
 \def\CMP{\em Commun. Math. Phys.}
\def\PRL{\em Phys. Rev. Lett.}

\def\MPL{\em Mod. Phys. Lett.}

\begin{document}
 \title[]
{Influence  of general convective motions on the exterior 
of 
isol\-ated rotating bodies in equilibrium}


\author{Ra\"ul Vera}
\address{School of Mathematical Sciences,
Queen Mary, University of London,
Mile End Road, London E1 4NS, UK}

\begin{abstract}
The problem of describing
isolated rotating bodies in equilibrium 
in General Relativity
has so far been treated under the 
assumption of
the circularity condition in the interior of the body.
For a fluid without energy flux, this condition implies
that the fluid flow moves only along the angular direction,
i.e. there is no convection.
Using this simplification, some recent studies have
provided us with uniqueness and existence results
for asymptotically flat vacuum exterior fields given
the interior sources.
Here, the generalisation of the problem to include general sources is
studied. It is proven that the convective motions
have no direct influence on the exterior field,
and hence, that the aforementioned results on uniqueness and existence
of exterior fields
apply equally in the general case.

\end{abstract}




\section{Introduction}
The theory for astrophysical self-gravitating isolated, rotating bodies in
equilibrium is still poorly understood in General Relativity (GR).
In particular, there is no known
explicit global model describing
the interior of a rotating object of a finite size in
stationary regime (equilibrium) together with
its exterior.
This fact has driven important efforts onto the finding of
theoretical general results for global models suitable
for those purposes.
Apart from stationarity, and since
we are interested in rotating objects, axial symmetry is a natural
assumption, and thus we will be interested in
spacetimes admitting a global two-dimensional
Abelian group
of isometries ($G_2$)  \cite{carter70} 
acting on timelike surfaces.
In addition, and to account for the isolation
of the body, the model is also required to be asymptotically
flat.

{}From the theoretical point of view, the most usual way to attack the global
problem consists in dividing the spacetime into two regions,
an {\em interior} region $\spI$, devised to describe our
spatially compact object, and a vacuum and
asymptotically flat {\em exterior} region $\spE$.
Both regions
are separated by the history of the limiting
surface of the body $\Sigma$.
Then, 
corresponding
Einstein field equations apply at
the interior and exterior sides taking common boundary data
at 
$\Sigma$.
In short, the global problem is divided into two,
namely the interior and the exterior problems.
They can be treated
independently by looking for existence and uniqueness
results on the spacetimes with boundary
$\spI$ and $\spE$
for given data on their respective boundaries $\Sigma^I$ and $\Sigma^E$.
Eventually, the two problems have to be
`matched' so that the two boundaries $\Sigma^I$ and $\Sigma^E$
can be identified as $\Sigma$.
Therefore, one has to look for compatible data on $\Sigma^I$ and $\Sigma^E$ 
through the imposition of the matching conditions.

Regarding the exterior problem, existence and uniqueness for
a stationary axially symmetric asymptotically flat vacuum solution
for a given interior
are under intense current investigation \cite{MASEuni,MarcERE,Marc?}\footnote{
For studies on the local problem see also \cite{LICH1} and \cite{hausernst}
and references therein. For accounts on the two problems separately
the reader is referred to \cite{shaupfis} and \cite{ansorg1},
references therein and references in \cite{MASEuni}.}.
Under the simplifying assumption of non-convective motions
in the interior region 
(see below),
uniqueness was solved in \cite{MASEuni} and
necessary conditions on the interior
for the existence of the exterior
have been already found \cite{Marc?}. These conditions
turn out to be also sufficient for static exteriors, 
and thus it has been conjectured
to be also true for the stationary ones \cite{Marc?}.
This issue is still under investigation.

A well known fact is that
the Abelian $G_2$ group on the vacuum exterior must act orthogonally
transitively \cite{papapetrou}.
In the works on global models for isolated rotating bodies in equilibrium,
and specifically those just mentioned,
a $G_2$ orthogonally transitive (OT)
has been always assumed
{\em also} on the interior region.
Denoting by $\vec\xi^I$ and $\vec\eta^I$ two independent vector fields
generating the $G_2$ at the interior, orthogonal transitivity is equivalent to
\begin{equation}
  \label{eq:OT}
  \star(\bm \xi^I \wedge \bm \eta^I \wedge \d \bm \eta^I)=
\star(\bm \xi^I \wedge \bm \eta^I \wedge \d \bm \xi^I)=0,
\end{equation}
where the star stands for the Hodge dual.
This is the so-called circularity condition,
and is equivalent to the absence of convective motions
in fluids without energy flux \cite{OT}.
Abusing the terminology, in the following
non-convective will refer to OT and vice versa.
This is, of course, a very restrictive condition,
since astrophysical objects 
are likely to naturally include non-circular motions.
In fact, recent studies on relativistic stars such as magnetars
already take into account the existence of toroidal fields or
meridional flows \cite{noconv2} (and references therein).

It has been recently shown \cite{mps} that
since the $G_2$ group is preserved through the matching
and it is OT in the exterior, (\ref{eq:OT}) will necessarily hold
on the matching hypersurface $\sup$ (see (\ref{eq:otsup}) below).
Once this result was proved, the remaining question
was whether (\ref{eq:otsup}) exhausts all the restrictions
on the $G_2$ group in the interior or not.

Herein, the role of convective motions 
is studied by performing the matching with the most general
interiors allowed by the present problem.
The result presented here shows how the condition found in \cite{mps}
is the only extra condition that appears
when generalising to a non-OT $G_2$ on the interior region,
and hence, how the non-OT terms (convective components
of the motions, in particular)
have no direct
effect on the exterior field nor on the external
shape of the body.
Clearly, this exhibits non-uniqueness of isolated
rotating bodies generating the same exterior field, under arbitrary
addition of convective components that vanish on the boundary of the body.
On the other hand,
this makes the aforementioned results \cite{MASEuni,MarcERE,Marc?}
applicable also in the general case.

It must be stressed that, of course, any assumption on the matter
model would relate convective components
to other quantities and hence, in an indirect manner,
eventually affect the shape of the body and the exterior field.
Note that no consideration on the matter model is made in this work.

\section{The matching conditions for a general interior}
When (\ref{eq:OT}) is not imposed on the interior region $\spI$,
allowing then for convective interiors,
there exists a coordinate system $\{T,\Phi,r,\zeta\}$ in which
the line-element reads \cite{WW1}
\be
\fl \rmd s^2_I=-e^{2V}\left(\rmd T+ B \rmd \Phi +
\ww ~ \rmd\zeta\right)^2+e^{-2V}
\left[e^{2h}\left(\rmd r^2+\rmd \zeta^2\right)+
\alpha^2\left(\rmd\Phi+\www ~ \rmd\zeta\right)^2\right],
\label{eq:ds2i}
\ee
where $V$, $B$, $h$, $\ww$, $\www$ and $\alpha$ are functions
of $r$ and $\zeta$ and $\vec\eta^I=\partial_\Phi$ is the axial Killing vector.
The stationary Killing
can be chosen to be ${\vec \xi}^I =\partial_T$. Conditions
(\ref{eq:OT}) are explicitly given by
\be
{\www}_{,r}=0,~~~~~
\left(B\www-\ww\right)_{,r}=0.
\label{eq:otexpl}
\ee
Under these conditions a coordinate change exists
which leaves the Killing vectors $\vec\xi^I$ and $\vec\eta^I$
invariant and sets $\www=\ww=0$ in (\ref{eq:ds2i}).
The metric for the exterior {\em vacuum} region $\spE$,
assumed to be free of ergoregions and/or Killing horizons,
can always be cast in the following form using the so-called Weyl coordinates
\be
\rmd s^2_E=-e^{2U}\left(\rmd t+ A \rmd\phi\right)^2+e^{-2U}
\left[e^{2k}\left(\rmd\rho^2+\rmd z^2\right)+
\rho^2 \rmd\phi^2\right],
\label{eq:ds2e}
\ee
where $U$, $A$, $k$ are functions
of $\rho$ and $z$. The axial Killing vector
is given by $\vec\eta^E=\partial_\phi$ and the axis of symmetry
is located at $\rho=0$.
The coordinate $t$ can be chosen to have an intrinsic meaning, namely to
measure proper time of an observer at infinity, and hence
the Killing vector $\vec\xi^E= \partial_t$ is unit at infinity.
With this choice, the
remaining coordinate freedom in (\ref{eq:ds2e}) consists only of
constant shifts of $t$, $\phi$ and $z$.

The matching preserving the stationarity and axial symmetry
is performed then as in \cite{MASEuni} (see also \cite{mps}),
to which the reader is referred to for details.
Local coordinates $\{\tau,\varphi,\lambda\}$
can be chosen in an abstract 
hypersurface $\suup$
such that the embedding $\embed_E: \suup\to \mm2$ is given by
\[
\embed_E:\{ t=\tau,\phi=\varphi,\rho=\rho(\lambda),z=z(\lambda)\},
\]
so that the images of $\partial_\tau$ and $\partial_\varphi$
by $\d \embed_E$ correspond to
$\partial_t|_{\embed_E(\suup)} (\equiv \vec e^E_{1})$
and $\partial_\varphi|_{\embed_E(\suup)} (\equiv \vec e^E_{2})$
respectively.
The image of the third vector $\partial_\lambda$, namely
$\vec e^E_{3}\equiv\dot\rho\partial_\rho+\dot z\partial_z|_{\embed_E(\suup)}$
where the dot indicates derivative with
respect to $\lambda$,
has been chosen orthogonal to $\vec e^E_{1}$ and $\vec e^E_{2}$.
Regarding the embedding for the interior,
the fact that the axial symmetry has an intrinsic meaning
imposes
$\d \embed_I(\partial_\varphi|_{\suup})=\partial_\Phi|_{\embed_I(\suup)}
(\equiv  \vec e^I_{2})$.
At this point the symmetry-preserving
matching introduces two parameters, $a$ and $b$,
by allowing
\[
\d\embed_I(\partial_\tau|_{\suup})=
a(\partial_T+b\partial_\Phi)|_{\embed_I(\suup)}.
\]
The linear coordinate change in $\spI$
\be
\Phi=\Phi'+a b T',~~
T=a T',
\label{eq:coc}
\ee
which implies $\partial_T'=a(\partial_T+b\partial_\Phi)$ and
$\partial_\Phi'=\partial_\Phi$,
is useful to deal with this freedom, since
it keeps (\ref{eq:ds2i}) (with primes)
and leaves invariant the axial Killing vector.
Substituting unprimed by primed quantities in (\ref{eq:ds2i}),
the new metric functions read
\begin{eqnarray}
&&\fl \alpha'=a \alpha,~~
h'-V'=h-V,\label{eq:cf1}\\
&&\fl e^{2V'}=a^2\left[(1+b B)^2 e^{2V}-\alpha^2 b^2 e^{-2V}\right],~~
B'=\frac{B(1+b B)e^{2V}-\alpha^2 b e^{-2V}}
{a\left[(1+b B)^2 e^{2V}-\alpha^2 b^2 e^{-2V}\right]},\label{eq:cf2}\\
&&\fl
W'_2=\frac{W_2 (1+b B)e^{2V}-\alpha^2 b W_3 e^{-2V}}
{a\left[(1+b B)^2 e^{2V}-\alpha^2 b^2 e^{-2V}\right]},~~
W'_3=W_3(1+b B)- b W_2.
\label{eq:cf3}
\end{eqnarray}
The functions $W_2$ and $W_3$ only appear in $W'_2$ and $W'_3$,
and therefore the expressions for $\alpha'$, $h'$, $V'$ and $B'$
in the OT case are exactly these very ones (see \cite{MASEuni}).

Now one can use the new
coordinate system (\ref{eq:coc}) by dropping primes
everywhere.
Nevertheless, it must be noticed
that if the interior $(\mm2_I,\g_I)$ is explicitly given,
then the freedom introduced by
the matching through (\ref{eq:cf1})-(\ref{eq:cf3}) 
must be taken into account.
In these new coordinates we obtain
$\d \embed_I(\partial_\tau|_{\suup})=\partial_T|_{\embed_I(\suup)}
(\equiv  \vec e^I_{1})$ together with
$\d \embed_I(\partial_\varphi|_{\suup})=\partial_\Phi|_{\embed_I(\suup)}
(\equiv  \vec e^I_{2})$ \cite{MASEuni}.
The continuity of the first fundamental form forces 
the image of $\partial_\lambda$, namely $\vec e^I_{3}$,
to be orthogonal
to $\vec e^I_{1}$ and $\vec e^I_{2}$, so that
$\vec e^I_{3}=\dot r \partial_r+\dot\zeta\left[(B \www -\ww)\partial_T-
\www \partial_\Phi + \partial_\zeta\right]|_{\embed_I(\suup)}$.
Therefore,
the most general form of the embedding
$\chi_I:\suup \to \mm2_I$ reads
\[
  \chi_I:\{T=\tau+f_T(\lambda),
\Phi=\varphi+f_\Phi(\lambda),r=r(\lambda),\zeta=\zeta(\lambda)\},
\]
where $\dot f_T=(B \www -\ww)|_{\suup}\dot \zeta$ and
$\dot f_\Phi=-\www|_{\suup} \dot \zeta$.
The four functions $\rho(\lambda)$,$z(\lambda)$, $r(\lambda)$
and $\zeta(\lambda)$ define then the matching hypersurface
$\sup\equiv \embed_I(\suup)=\embed_E(\suup)$
and will be determined, as shown below, by the matching conditions.
The explicit expressions of the (non-unit)\footnote{The only necessary
requirement for the matching is that they have the same norm and
relative orientations.} normal forms
to the matching hypersurface are
${\bm n}^I=e^{2h}(\dot\zeta \rmd r + \dot r \rmd \zeta)|_{\sup}$
and
${\bm n}^E=e^{2k}(\dot z \rmd\rho + \dot\rho \rmd z)|_{\sup}$.

In the OT case, once the interior
metric $\g_I$ is known,
the whole matching conditions were shown in \cite{merce,MASEuni}
to reorganise into
three sets of conditions: namely, (a) a set of conditions on the interior
hypersurface, (b) a set of conditions for the exterior hypersurface,
and (c) boundary conditions for the exterior problem.

In the present general case the three sets are recovered
again. Indeed, the whole
set of matching conditions can be cast in the following form:
\\
(a) {\em Conditions on the interior hypersurface:}
\begin{eqnarray}
  \label{eq:conda1}
 &&n^{I \alpha} n^{I \beta} S_{\alpha\beta}|_{\sup} =0,~~~
 n^{I \alpha} e^{I \beta}_3 S_{\alpha\beta}|_{\sup} =0,\\
  \label{eq:conda2}
&& {\www}_{,r}|_{\sup} =0,~~~
\left(B\www-\ww\right)_{,r}|_{\sup} =0,
\end{eqnarray}
where $S_{\alpha\beta}$ is the Einstein tensor in the interior region.
The two new equations (\ref{eq:conda2})
are equivalent to (see (\ref{eq:otexpl}))
\begin{equation}
  \label{eq:otsup}
\star(\bm \xi^I \wedge \bm \eta^I \wedge \d \bm \eta^I)|_{\sup} =
\star(\bm \xi^I \wedge \bm \eta^I \wedge \d \bm \xi^I)|_{\sup}=0,
\end{equation}
as expected,
since they were already shown in \cite{mps}
to be necessary conditions (even in more general settings).
Importantly, decomposing the Einstein tensor at the interior
into its OT and non-OT part, so that
$S_{\alpha\beta}=S_{\alpha\beta}^{OT} + S^W_{\alpha\beta}$ defining
$S_{\alpha\beta}^{OT} \equiv
S_{\alpha\beta}|_{\xi^I \wedge \eta^I \wedge d \eta^I=
\xi^I \wedge \eta^I \wedge d \xi^I=0}$,
then equations (\ref{eq:conda1}) are equivalent to
\begin{equation}
  \label{eq:conda1OT}
   n^{I \alpha} n^{I \beta} S^{OT}_{\alpha\beta}|_{\sup} =0,~~~
 n^{I \alpha} e^{I \beta}_3 S^{OT}_{\alpha\beta}|_{\sup} =0,
\end{equation}
provided that (\ref{eq:conda2}) is satisfied. 
Although the decomposition is not invariantly defined,
the contractions appearing in equations (\ref{eq:conda1OT}) are.\footnote{As
a matter of fact,
one always has
$n^{I \alpha} e^{I \beta}_3 S^{W}_{\alpha\beta}|_{\sup}=0 $
and
$n^{I \alpha} n^{I \beta} S^{W}_{\alpha\beta}|_{\sup}=
\left.\frac{\vec e_3^2\,\, \vec\xi^2}
{4\left[(\vec \xi \cdot \vec \eta)^2- \vec \xi^2 \vec \eta^2\right]^2}
\left[\star(\bm\xi^I \wedge \bm\eta^I \wedge \d \bm\xi^I)\vec\eta
+\star(\bm\xi^I \wedge \bm\eta^I \wedge \d \bm\eta^I)\vec\xi\right]^2
\right|_{\sup},
$
with the obvious notation $\vec v^2\equiv v^\alpha v_\alpha$.}
Therefore, the system composed by
(\ref{eq:conda1}) and (\ref{eq:conda2})
is equivalent to (\ref{eq:conda1OT}) and (\ref{eq:conda2}).
Equations (\ref{eq:conda1OT}) constitute the set (a) in \cite{MASEuni}.

Equations (\ref{eq:conda1})-(\ref{eq:conda2})
form an overdetermined system of four ordinary differential
equations for $r(\lambda)$ and $\zeta(\lambda)$.
If a solution exists, then the matching will be possible.
Generically, the interior matching hypersurface will be
uniquely determined. Nevertheless, there are cases where
(\ref{eq:conda1})-(\ref{eq:conda2}) contain no information,
and hence the matching is possible across any timelike
hypersurface preserving the symmetry.

As mentioned, the difference between the OT 
and the 
non-OT cases lies in
the fulfilment of equations (\ref{eq:otexpl}).
Conditions (\ref{eq:conda2}) state
that (\ref{eq:otexpl}) hold on the matching hypersurface.
At this point 
only the normal derivatives of (\ref{eq:otexpl})
could make a difference with respect to the OT case
in the rest of the matching conditions.
Surprisingly, after a straightforward
calculation, these normal derivatives can be shown not
to appear.
Indeed, the rest of the matching conditions are exactly the same
as in the OT case \cite{MASEuni}. These are
the following:
\\
(b) {\em Exterior matching hypersurface:}
Once the interior hypersurface has been determined,
the functions defining the exterior matching hypersurface
$\rho(\lambda)$ and $z(\lambda)$ are uniquely determined by
\begin{equation}
  \label{eq:condb}
  \rho(\lambda)=\alpha|_{\sup},\hspace{1cm}
  \dot z(\lambda)=\alpha_{,r} \dot \zeta-\alpha_{,\zeta}\dot r |_{\sup}.
\end{equation}
Note that an additive constant in $z(\lambda)$ is spurious
due to the freedom $z\to z+const$.
\\
(c) {\em Boundary conditions for the exterior problem:}
The rest of the matching conditions read
\begin{equation}
  \label{eq:condc}
  \begin{array}[]{cc}
U|_{\sup}=V|_{\sup},&A|_{\sup}=B|_{\sup},\\
\vec n^E (U)|_{\sup}=\vec n^I (V)|_{\sup},&
\vec n^E (A)|_{\sup}=\vec n^I (B)|_{\sup}.
  \end{array}
\end{equation}
These four equations provide the boundary conditions
on the metric functions $U$ and $A$,
which translate into overdetermined boundary conditions
on the Ernst potential for the exterior vacuum problem.
To be more precise, the boundary conditions
leave still a degree of freedom in the form of an additive constant
in the twist potential (see \cite{MASEuni}).
Nevertheless, as shown in \cite{MASEuni}, if a solution exists, then
this additive constant is fixed and thus the Ernst
potential
is determined everywhere in the exterior region.
The remaining function $k$ in (\ref{eq:ds2e})
is found, up to an additive constant, by quadratures.
This constant is fixed
by the matching procedure by using
the complementary equation
\begin{equation}
\label{eq:condk}
k|_{\sup}=
\left.\left[h-\frac{1}{2}\ln\left(\alpha^2_{,r}+\alpha^2_{,\zeta}\right)
\right]\right|_{\sup}.
\end{equation}
This matching condition is complementary in the sense that
the previous equations (\ref{eq:conda1})-(\ref{eq:conda2})
ensure that its derivative with respect to $\lambda$
is satisfied, and hence (\ref{eq:condk})
only determines the additive constant in $k$.
This comment was incidentally left out in \cite{MASEuni}.

\subsection{Rewriting the set (a)}
\label{subsec}
Conditions (\ref{eq:otsup}) are purely geometrical.
In \cite{merce}, two matching conditions
were rewritten
in terms of the Einstein tensor (so-called Israel conditions \cite{ISRA})
and the analogous procedure has been used here 
to get (\ref{eq:conda1}).
The question that arises is whether (\ref{eq:conda2})
are also equivalent to any of the Israel conditions,
since that would relate the geometrical
aspect of (\ref{eq:otsup}) with physical properties
of the interior matter content on $\sup$. 
Let me stress here the fact that
the Israel conditions are consequences of the matching conditions,
and hence, necessary (not sufficient in general)
conditions for the matching.
Since we have a vacuum exterior, the Israel conditions read
\begin{equation}
  \label{eq:israel}
  \begin{array}[c]{ll}
(\mbox{i})~~~ n^{I \alpha} n^{I \beta} S_{\alpha\beta}~|_{\sup} =0,&
(\mbox{ii})~~~ n^{I \alpha} e^{I \beta}_3 S_{\alpha\beta}~|_{\sup} =0.\\
(\mbox{iii})~~~ n^{I \alpha} e^{I \beta}_1 S_{\alpha\beta}~|_{\sup} =0,&
(\mbox{iv})~~~ n^{I \alpha} e^{I \beta}_2 S_{\alpha\beta}~|_{\sup} =0.
  \end{array}
\end{equation}
In the OT case the two last relations
(iii) and (iv)
are identically satisfied
because of the structure of the
Einstein tensor inherited by the OT $G_2$.
Therefore, in the OT case the two equations (\ref{eq:conda1})
constitute the whole set of non-trivial Israel conditions, and thus
the set (a) of matching conditions is made up by the Israel conditions.
But in the general case the two last relations in (\ref{eq:israel})
are non-trivial
(of course, they will be satisfied
once the whole set of matching conditions
(\ref{eq:conda1}),(\ref{eq:conda2}),(\ref{eq:condb}),(\ref{eq:condc})
is satisfied).

With the help of a  well known identity \cite{papapetrou,OT}
one can find
\begin{equation}
  \label{eq:final}
  \begin{array}[c]{r}
\displaystyle{n^{I \alpha} e^{I \beta}_1 S_{\alpha\beta}~|_{\sup} =
    -\left.\frac{{e^{2V}}}{2\alpha}\right|_{\sup}
\frac{\rmd}{\rmd \lambda}\left[\left.
\star(\bm \xi^I \wedge \bm \eta^I \wedge \d \bm \xi^I)\right|_{\sup}\right]},
\\
\displaystyle{ n^{I \alpha} e^{I \beta}_2 S_{\alpha\beta}~|_{\sup} =
    -\left.\frac{{e^{2V}}}{2\alpha}\right|_{\sup}
\frac{\rmd}{\rmd \lambda}\left[\left.
\star(\bm \xi^I \wedge \bm \eta^I \wedge \d \bm \eta^I)\right|_{\sup}\right]},
  \end{array}
\end{equation}
on the matching hypersurface.
{}From these identities it readily follows that
(\ref{eq:otsup})
imply the Israel conditions (iii) and (iv), as mentioned.
On the other hand, and more interestingly, if 
(iii) and (iv) hold, then
$\star(\bm \xi^I \wedge \bm \eta^I \wedge \d \bm \xi^I)|_{\sup}$ and
$\star(\bm \xi^I \wedge \bm \eta^I \wedge \d \bm \eta^I)|_{\sup}$
are constants on 
$\sup$.
Now, if our interior region is to describe a spatially compact
and simply connected object $\sup$ will intersect the axis of symmetry.
At those points
$\bm \eta^I$ vanishes, and thus
$\star(\bm \xi^I \wedge \bm \eta^I \wedge \d \bm \xi^I)|_{\sup}$ and
$\star(\bm \xi^I \wedge \bm \eta^I \wedge \d \bm \eta^I)|_{\sup}$
will vanish.
This argument involving the axis
is analogous to that used to show that
the exterior region must admit a OT $G_2$ \cite{papapetrou,OT}.
Therefore, in the cases we will be interested in, the Israel
conditions (iii) and (iv) are equivalent to the conditions
(\ref{eq:otsup}). The relation between
the geometrical properties of the $G_2$ at the interior
with the properties of the matter content on the boundary of the body
is then manifest.

\section{Summary and conclusions}
The complete set of matching conditions for the general case
can be cast
into the three sets (a), (b) and (c), as described
above, in analogy with the OT case.
In fact, the only difference with the OT case
lies in the set (a), where now we have two more equations (\ref{eq:otsup}).
Notice also that although there might be hidden non-OT terms 
in (\ref{eq:conda1}), this is not the case, as (\ref{eq:conda1})
are equivalent to (\ref{eq:conda1OT}).

In subsection \ref{subsec} it has been shown that 
for spatially compact and simply connected interiors
(\ref{eq:conda1})-(\ref{eq:conda2})
are equivalent to (\ref{eq:israel}),
and therefore the set (a) of conditions is equivalent to
the complete set of Israel conditions. 
This is analogous to what happens
in the non-convective case, the difference being that
the Israel conditions in the present general case
constitute four relations instead of only two.

The rest of the conditions (sets (b) and (c)), which concern
the unknown exterior region, are the same as in the OT case.
This fact is important, because
it demonstrates that the exterior problem
is ``independent'' of any convective motions in the interior.
 More precisely,
 once
 the non-OT terms
 have been proven to vanish
 on the boundary of the body, see (\ref{eq:otsup}),
 there is no other explicit information
 coming from the inner non-OT terms
 affecting the boundary conditions
 (nor the exterior boundary itself)
 on the Ernst potential for the exterior problem.
Therefore,
all the existing results and studies on the existence and
uniqueness of asymptotically flat vacuum exteriors
such as \cite{merce,MASEuni,MarcERE,Marc?},
where the circularity condition has been always assumed, 
apply equally in the general case.

In particular, the existence of an asymptotically flat
vacuum exterior has been shown to impose a number of conditions
on the overdetermined boundary data
for the Ernst potential \cite{Marc?}.
Once we have an interior region ``shaped'' by the set (a),
those conditions translate onto the
interior quantities through (\ref{eq:condc})
together with (\ref{eq:condb}), 
becoming the conditions
our interior 
has to satisfy in order to describe a truly isolated body.
The point made here implies
that the existence of an asymptotically flat vacuum exterior 
poses no conditions onto the possible convective
motions inside the body.

On the other hand,
we know that the exterior field generated by
a non-convective interior region 
describing an isolated  rotating body in equilibrium
is unique\footnote{Once the identification
of the interior with the exterior through $\sup$ has been prescribed
by fixing $a$ and $b$.} \cite{MASEuni}.
Now, given a global model composed
by an OT interior together with
its corresponding asymptotically flat exterior, 
so that (\ref{eq:conda1OT}),
(\ref{eq:condb}), (\ref{eq:condc}) (and (\ref{eq:condk})) hold on,
say, $\sup_{OT}$,
one can always introduce arbitrary convective components
vanishing on $\sup_{OT}$ and thus generate different interiors
keeping the same shape and exterior fields.
Conversely, given a general interior explicit
metric $g_I$ such that the set (a)
of conditions is satisfied for some hypersurface $\sup^I$,
the exterior (if it exists) will be unique and 
the same as the one generated by $g_I|_{W_2=W_3=0}$.

\section*{Acknowledgements}
I am grateful to Malcolm MacCallum, Marc Mars, Jos\'e Senovilla and
Reza Tavakol for their careful readings, comments and valuable suggestions
that have led to very significant improvements to previous versions
of the manuscript. 
I also thank the EPSRC for grant no. MTH 03 R AJC6.

\end{document}